\documentclass[11pt]{article}
\usepackage{amsmath}
\usepackage{amssymb}
\usepackage{amsthm,amsxtra}
\usepackage{epsf}
\usepackage{eepic}
\newlength{\mytopmargin}
\newlength{\myleftmargin}
\newtheorem{lemma}{Lemma}
\newtheorem{prop}{Proposition}
\newtheorem{thm}{Theorem}
\newtheorem{cor}{Corollary}

\begin{document}

\title{\Large\bf A random matrix decimation procedure relating 
$\beta = 2/(r+1)$ to $\beta = 2(r+1)$}
\author{Peter J. Forrester}
\date{}
\maketitle

\begin{center}
\it
 $^{\dagger}$Department of Mathematics and Statistics, University of Melbourne, \\
Victoria 3010, Australia
\end{center}

\bigskip
\begin{center}
\bf Abstract
\end{center}
\par
\bigskip
\noindent
Classical random matrix ensembles with orthogonal symmetry have the property that the
joint distribution of every second eigenvalue is equal to that of a classical random matrix
ensemble with symplectic symmetry. These results are shown to be the case $r=1$ of a
family of inter-relations between eigenvalue probability density functions for
generalizations of the classical random matrix ensembles referred to as
$\beta$-ensembles. The inter-relations give that the joint distribution
of every $(r+1)$-st eigenvalue in certain $\beta$-ensembles with $\beta = 2/(r+1)$ is
equal to that of another $\beta$-ensemble with $\beta = 2(r+1)$. The proof requires 
generalizing a conditional probability density function due to Dixon and Anderson.

\newpage
\section{Introduction}
\subsection{The setting and summary of results}
The Dixon-Anderson conditional probability density function (PDF) refers to the function of
$\{\lambda_j\}$ specified by \cite{Di05, An91}
\begin{equation}\label{d1}
{\Gamma(\sum_{j=1}^n s_j) \over \Gamma(s_1) \cdots \Gamma(s_n)}
{\prod_{1 \le j < k \le n - 1} (\lambda_j - \lambda_k) \over \prod_{1 \le j < k \le n} (a_j - a_k)^{s_j+s_k-1} }
\prod_{j=1}^{n-1} \prod_{p=1}^n | \lambda_j - a_p |^{s_p - 1} \chi_A,
\end{equation}
where generally $\chi_A = 1$ if condition $A$ is true, $\chi_A = 0$ otherwise, and 
here condition
$A$ is the inequalities
$$
a_1 > \lambda_1 > a_2 > \lambda_2 > \cdots > \lambda_{n-1} > a_n
$$
specifying an interlaced region.
An analogous conditional PDF for angles $\{\psi_j\}_{j=1,\dots,n}$, due to Forrester and Rains
\cite{FR05}, is specified by
\begin{eqnarray}\label{d2}
{\Gamma^2((\sum_{j=0}^{n-1}\alpha_j + 1)/2) \over 2 \pi \Gamma(\alpha_0) \cdots \Gamma(\alpha_{n-1}) }
{\prod_{1 \le j < k \le n} | e^{i \psi_k} - e^{i \psi_j} | \over
\prod_{1 \le j < k \le n}
|e^{i \theta_k} - e^{i \theta_j} |^{\alpha_j + \alpha_k - 1} } 
\prod_{j=1}^{n} \prod_{p=1}^n | e^{i \psi_j} - e^{i \theta_p} |^{\alpha_j - 1} \chi_R
\end{eqnarray}
where $\theta_n = 2\pi$ and $R$ denotes the interlaced region
\begin{equation}\label{2.1}
0 < \psi_1 < \theta_1 < \psi_2 <  \theta_2 < \cdots < \psi_n < 2 \pi
\end{equation}
(in fact in \cite{FR05} the parameters were specialized to $\alpha_1 = \cdots = \alpha_{n-1} = \alpha$, 
but the working therein applies without this specialization). The conditional PDF (\ref{d1}) is fundamental in the
theory of the Selberg integral \cite[Ch.~3]{Fo02}, \cite{FW07}. 
It permits various generalizations, which are intimately related to
Macdonald polynomial theory \cite{Ok97} and the theory of multivariable elliptic hypergeometric functions
\cite{Ra06}. 

It is the purpose of this paper to introduce different generalizations of
(\ref{d1}) and (\ref{d2}), and to discuss their consequences to random matrix theory.
To state our generalizations, which will be derived in Section 2, we require for the
normalizations the Selberg integral \cite{Se44}
\begin{eqnarray}
S_N(\lambda_1,\lambda_2,\lambda) & := &
\int_0^1 dt_1 \cdots \int_0^1 dt_N \,
 \prod_{l=1}^N t_l^{\lambda_1}(1-t_l)^{\lambda_2}
 \prod_{1 \le j < k \le N} |t_k - t_j|^{2 \lambda} \nonumber \\
& = & 
 \prod_{j=0}^{N-1} {\Gamma (\lambda_1 + 1 + j\lambda)
\Gamma (\lambda_2 + 1 + j\lambda)\Gamma(1+(j+1)\lambda) \over
\Gamma (\lambda_1 + \lambda_2 + 2 + (N + j-1)\lambda) \Gamma (1 + \lambda )},
\label{3.1}
\end{eqnarray}
and the Morris type integral (see \cite[Ch.~3]{Fo02}) 
\begin{eqnarray}
M_N(a,\lambda) & := & (2 \pi)^N
\int_{-1/2}^{1/2} d \theta_1 \cdots \int_{-1/2}^{1/2} d \theta_N \,
\prod_{l=1}^N |1 + e^{i \theta_l} |^{2a} 
\prod_{1 \le j < k \le N} | e^{ i \theta_k} - e^{ i \theta_j} |^{2 \lambda} 
\nonumber \\
& = & (2 \pi)^N \prod_{j=0}^{N-1}
{\Gamma(\lambda j + 2a + 1) \Gamma(\lambda(j+1) + 1) \over
(\Gamma(\lambda j + a + 1) )^2 \Gamma(1 + \lambda) }.
\label{3.4ab}
\end{eqnarray}

\begin{thm}\label{tt1}
Let $r \in \mathbb Z^+$.
The Dixon-Anderson PDF (\ref{d1}) is the $r=1$ case of the family of conditional
PDFs
\begin{equation}\label{d3}
{1 \over \hat{C} }
{
\prod_{1 \le j < k \le r(n - 1)} (\lambda_j - \lambda_k)^{2/(r+1)} \over 
\prod_{1 \le j < k \le n} (a_j - a_k)^{r(s_j+s_k -2/(r+1))} }
\prod_{j=1}^{r(n-1)} \prod_{p=1}^n | \lambda_j - a_p |^{s_p - 1} \chi_{A_r}
\end{equation}
where $A_r$ is the interlaced region
\begin{equation}\label{1.6a}
a_j > \lambda_{r(j-1) + 1} > \lambda_{r(j-1) + 2} > \cdots  > 
\lambda_{r(j-1) + r - 1} > a_{j+1} \qquad (j=1,\dots,n-1).
\end{equation}
and the normalization $\hat{C}$ is specified by
\begin{equation}\label{d3a}
\hat{C} = 
\prod_{l=1}^{n-1} {1 \over r!} S_r \Big (  \sum_{p=1}^l s_p + 2(l-1)r/(r+1) - l, s_l - 1,
1/(r+1) \Big ).
\end{equation}
The circular analogue of the Dixon-Anderson PDF (\ref{d2}) is the case $r=1$ of the
family of conditional PDFs
\begin{equation}\label{d4}
{1 \over \tilde{C} }
{ \prod_{1 \le j < k \le rn} 
| e^{i \psi_k} - e^{i \psi_j} |^{2/(r+1)}  \over
\prod_{1 \le j < k \le n}
|e^{i \theta_k} - e^{i \theta_j} |^{r(\alpha_j + \alpha_k -2/(r+1))} } 
\prod_{j=1}^{rn} \prod_{p=1}^n | e^{i \psi_j} - e^{i \theta_p} |^{\alpha_j - 1} \chi_{R_r},
\end{equation}
where $ \theta_n = 2 \pi$, $R_r$ denotes the interlaced region
\begin{equation}\label{1.8a}
\theta_{j-1} < \psi_{(r-1)j+1} < \psi_{(r-1)j+2} < \cdots < \psi_{rj} < \theta_j \quad (j=1,\dots,n)
\end{equation}
with $\theta_0 := 0$, and the normalization $\tilde{C}$ is specified by
\begin{equation}\label{d3a1}
\tilde{C} =  \hat{C}  |_{\{s_p\} \mapsto \{\alpha_p\}} 
M_r \Big ( {1 \over 2} \Big (\sum_{p=1}^n \alpha_p +
2(n-1) r/(r+1) - n \Big ), 1/(r+1)   \Big ).
\end{equation}
\end{thm}

As to be revised in Section 3, both (\ref{d1}) and (\ref{d2}) have consequences in random matrix
theory, in that they provide inter-relations between ensembles with orthogonal, unitary and
symplectic symmetry. Likewise, it will be shown in Section 4 that for general $r \in \mathbb Z^+$ the
generalizations (\ref{d3}) and (\ref{d4}) have consequences in random matrix theory. To
state these consequences, let ME${}_{\beta,N}(g)$ refer to the matrix ensemble specified by the
eigenvalue PDF
\begin{equation}\label{f.1}
{1 \over C} \prod_{l=1}^N g(x_l) \prod_{1 \le j < k \le N} |x_k - x_j|^\beta
\end{equation}
(unless otherwise stated, $C$ will denote {\it some} normalization), and let the
eigenvalues in (\ref{f.1}) be ordered
\begin{equation}\label{f.2}
x_1 > x_2 > \cdots > x_N.
\end{equation}
Also, let CE${}_{\beta,N}^{b}$ refer to the so called circular Jacobi $\beta$-ensemble of unitary
random matrices (see e.g.~\cite[Ch.~2]{Fo02}), which is specified by the eigenvalue PDF
\begin{equation}\label{f.3}
{1 \over C} \prod_{l=1}^N |1 - e^{i \theta_l} |^b \prod_{1 \le j < k \le N} |e^{i \theta_k} -
e^{i \theta_j} |^\beta,
\end{equation}
and suppose the angles are ordered
\begin{equation}\label{f.4}
0 < \theta_1 < \theta_2 < \cdots < \theta_N < 2 \pi.
\end{equation}
Denote by ${\rm D}_{p}({\rm M}{\rm E}_{\beta,N} (g) )$ the joint
marginal distribution of $x_p,x_{2p},x_{3p},\dots$ in (\ref{f.1}), and denote
by ${\rm D}_{p}({\rm C} {\rm E}_{\beta, N}^{b}  )$ the joint
marginal distribution of $\theta_p,\theta_{2p},\theta_{3p},\dots$ in (\ref{f.3}).
Following the nomenclature of \cite{FR02b}, the D here stands for a decimation procedure.

\begin{thm}\label{tt2}
Let $r \in \mathbb Z^+$. One has the inter-relations between
matrix ensembles
\begin{eqnarray}\label{tt3}
{\rm D}_{r+1}({\rm M} {\rm E}_{2/(r+1), (r+1)N+r} (x^a (1 - x)^b) ) & = &
{\rm M} {\rm E}_{2(r+1),N} (x^{(r+1)a+2r} (1 - x)^{(r+1)b+2r}) \nonumber \\ 
{\rm D}_{r+1}({\rm M} {\rm E}_{2/(r+1), (r+1)N} ((1 - x)^b) ) & = &
{\rm M} {\rm E}_{2(r+1),N} ((1 - x)^{(r+1)b+2r}) \nonumber \\ 
{\rm D}_{r+1}({\rm M} {\rm E}_{2/(r+1), (r+1)N+r} (x^a e^{-x}) ) & = &
{\rm M} {\rm E}_{2(r+1),N} (x^{(r+1)a+2r} e^{-(r+1)x} ) \nonumber \\ 
{\rm D}_{r+1}({\rm M} {\rm E}_{2/(r+1), (r+1)N} (e^{-x})) & = &
{\rm M} {\rm E}_{2(r+1),N} (e^{-(r+1)x}) \nonumber \\ 
{\rm D}_{r+1}({\rm M} {\rm E}_{2/(r+1), (r+1)N+r} (e^{-x^2}) ) & = &
{\rm M} {\rm E}_{2(r+1),N} (e^{-(r+1)x^2} ) \nonumber \\ 
{\rm D}_{r+1}({\rm CE}^b_{2/(r+1),(r+1)N+r}) & = &
{\rm CE}_{2(r+1),N}^{(r+1)b+2r} \nonumber \\
{\rm D}_{r+1}({\rm CE}^0_{2/(r+1),(r+1)N}) & = &
{\rm CE}_{2(r+1),N}^{0}.
\end{eqnarray}
\end{thm}

In Section 5 these inter-relations, applying at the level of joint eigenvalue
PDFs, will be shown to imply analogous inter-relations linking
$\beta = 2(r+1)$ and $\beta = 2/(r+1)$ for marginal distributions of single
eigenvalues. Depending on these single eigenvalues being in the neighbourhood
of the spectrum edge, or in the bulk, there are three distinct $N \to \infty$
scalings, implying corresponding inter-relations for marginal distributions
in the scaled ensembles.

\subsection{Significance of the $\beta$ ensembles}
One interpretation of (\ref{f.1}) is as the Boltzmann factor for a classical
gas at inverse temperature $\beta$ with potential energy
$$
- {1 \over \beta} \sum_{l=1}^N \log g(x_l) - \sum_{1 \le j < k \le N}
\log | x_k - x_j|.
$$
Because of the pairwise logarithmic replusion (two-dimensional Coulomb law), such
a classical gas is referred to as a log-gas. This interpretation allows for a
number of properties of correlations and distributions to be anticipated using
arguments based on macroscopic electrostatics. Of interest
for purposes of the present study is a prediction relating to the probability $E_\beta^{\rm bulk}(n;2t)$
that an interval of size $2t$ in the bulk contains exactly $n$ eigenvalues, in
the $N \to \infty$ limit of (\ref{f.1}),
scaled so the eigenvalue density is unity (by universality, for a large class of $g$
this limit is expected to be independent of $g$). The prediction is that
\cite{Dy95,FS95}
\begin{equation}\label{10.m2}
\log E_\beta^{\rm bulk} (n;2t) \mathop{\sim}\limits_{t,n \to \infty \atop t \gg n}
- \beta {(\pi  t)^2 \over 4} + \Big
( \beta n + {\beta \over 2} -1 \Big ) \pi  t
+ {n \over 2} \Big ( 1 - {\beta \over 2} - {\beta n \over 2} \Big )
\Big ( \log {8 \pi  t \over n} + 1 \Big ).
\end{equation}
We will see in Section \ref{ssp} that this is consistent with the final relation
in (\ref{tt3}).

A second interpretation of (\ref{f.1}) is as the absolute value squared of the
ground state wave function for certain quantum many body problems with inverse
square pair potentials (Calogero-Sutherland systems). For example, with
$g(x) = e^{- \beta x^2/2}$, (\ref{f.1}) is equal to $|\psi_0|^2$ where $\psi_0$
is the eigenfunction corresponding to the smallest eigenvalue of the
Schr\"odinger operator
$$
H  := - \sum_{j=1}^N {\partial^2 \over \partial x_j^2}
+{\beta^2 \over 4} \sum_{j=1}^N x_j^2 +\beta (\beta /2 - 1) \sum_{1 \le j < k \le N}
{1 \over (x_j - x_k)^2}.
$$
Such interpretations are restricted to a total of four examples of
(\ref{f.1}) or (\ref{f.3}) (see e.g.~\cite{Fo02}), namely
$g(x)$ equal to
$$
x^a(1-x)^b \quad (0<x<1), \qquad x^a e^{-x} \quad (x>0), \qquad e^{-x^2}
$$
in (\ref{f.1}) and $b=0$ in (\ref{f.3}).
In the study of these ground
states, identities of a different type to Theorem \ref{tt2} relating $\beta$
to $4/\beta$ ($\beta$ even) have previously been encountered. These are so called
duality relations, an example being \cite{BF97a}
$$
\Big \langle \prod_{j=1}^N (t - x_j)^m \Big \rangle_{{\rm ME}_{\beta,N}(e^{-\beta x^2/2})} =
\Big \langle \prod_{j=1}^m (t - i x_j)^N \Big \rangle_{{\rm ME}_{4/\beta,m}(e^{- x^2})}.
$$

The four examples of (\ref{f.1}) or (\ref{f.3}) which permit interpretations as the
ground state of quantum many body systems can also be realized as the eigenvalue
PDF for certain ensembles of random matrices. Thus, for example, with
$\tilde{\chi}_k$ denoting  value drawn from the square root of the gamma
distribution $\Gamma[k/2,1]$, and N$[0,1]$ denoting a number drawn from the
standard normal distribution, the tridiagonal matrix
$$
T_\beta := \left [ \begin{array}{ccccc} {\rm N}[0,1] &
\tilde{\chi}_{(N-1)\beta} & & & \\
\tilde{\chi}_{(N-1)\beta} & {\rm N}[0,1] & \tilde{\chi}_{(N-2)\beta} & & \\
 & \tilde{\chi}_{(N-2)\beta} & {\rm N}[0,1] & \tilde{\chi}_{(N-3)\beta} & \\
& \ddots & \ddots & \ddots & \\
&&\tilde{\chi}_{2\beta} & {\rm N}[0,1] & \tilde{\chi}_{\beta} \\
& & & \tilde{\chi}_{\beta} & {\rm N}[0,1] \end{array} \right ]
$$
has its eigenvalue PDF given by ME${}_{\beta,N}(e^{-x^2/2})$
\cite{DE02}. Taking various scaled $N \to \infty$ limits
(see Section \ref{ssp}) of these ensembles of random matrices leads to a 
description of the limiting eigenvalue distributions in terms of stochastic
differential operators \cite{ES06,RRV06,KS06}. Further, for eigenvalues in the bulk,
there is a description in terms of a process associated with Brownian motion
in the hyperbolic plane \cite{VV07}.

It is an open problem, for general $r$, to derive the results of Theorem
\ref{tt2} as consequences of the eigenvalue PDFs being realizable from
concrete random matrix ensembles. In the case $r=1$ such matrix theoretic
derivations have been given (excluding the sixth identity which relates to
CE${}_N^b$) using matrix realizations not applicable
for general $r$ \cite{FR02b,FR05}.

\section{Proof of Theorem \ref{tt1}}
\setcounter{equation}{0}
\subsection{Analytic continuation of $L_{r,n}(\{a_p\})$ --- a cancellation effect}
Let
\begin{eqnarray}
&& \! \! \! \! \!L_{r,n}(\{a_p\}) :=  \int_{A_r} d \lambda_1 \cdots d \lambda_{r(n-1)} \,
\prod_{1 \le j < k \le r(n-1)} (\lambda_j - \lambda_k)^{2/(r+1)}
\prod_{j=1}^{r(n-1)} \prod_{p=1}^n | \lambda_j - a_p|^{s_p - 1} \label{L1} \\
&& \! \! \! \! \!R_{r,n}(\{a_p\})  :=  \prod_{1 \le j < k \le n} (a_j - a_k)^{r(s_j + s_k - 
2/(r+1))}, \label{R1}
\end{eqnarray}
where in (\ref{L1}) $A_r$ is given by (\ref{1.6a}).
To prove that (\ref{d3}) is a PDF in $\{\lambda_j\}$ we must show that
(\ref{L1}) is proportional to (\ref{R1}). In preparation for this task, we will
study the analytic continuation of (\ref{L1}), and in particular its value
upon the interchange of the order of $\{a_p\}$. For the latter purpose,
it suffices to consider the effect of the interchange of an arbitrary
neighbouring pair of parameters $a_l,a_{l+1}$ say. One finds that, up to a
phase, the analytic continuation is a symmetric function in
$\{(s_p,a_p)\}$.

\begin{prop}\label{pr1}
Interchanging the order of the $\{a_p\}$ on the real line
via analytic continuation, and making a
corresponding change to the order of $\{s_p\}$, gives back the same integral
representation (\ref{L1}), up to a phase. In particular
\begin{equation}\label{Las}
L_{r,n}(\{a_p\}) \Big |_{a_l \leftrightarrow a_{l+1} \atop
s_l \leftrightarrow s_{l+1} } =
e^{- \pi i r (s_l + s_{l+1} - 2/(r+1) ) }
L_{r,n}(\{a_p\}).
\end{equation}
\end{prop}

The proof of Proposition \ref{pr1} relies crucially on a certain cancellation
effect, isolated by the following result.

\begin{lemma}\label{Lem1}
Consider an arrangement of $r$ $0's$ and $q$ $1's$ $(1 \le q \le r)$ in a
line. Let this be considered as the sequence ${\cal A} = (n_j)_{j=1,\dots,r+q}$
with each $n_j=0$ or 1. Further let
\begin{equation}\label{Kn}
K(n_j) = \left \{ \begin{array}{ll}0, & n_j = 0 \\
\#0{\rm '}s \: {\rm to \: the \: right \: of \:} n_j, & n_j = 1,
\end{array} \right.
\end{equation}
and use this to specify the statistic
\begin{equation}\label{2.4a}
K({\cal A}) = \sum_{j=1}^{r+q} K(n_j).
\end{equation}
One has
\begin{equation}\label{L1f}
\sum_{\cal A} e^{- 2 \pi i K({\cal A})/(r+1)} = 0.
\end{equation}
\end{lemma}

\noindent
Proof.  \quad The definition (\ref{Kn}) can be written
$$
K(n_j) = n_j \sum_{k=j+1}^{r+q} (1 - n_j),
$$
and this substituted in (\ref{2.4a}) gives
$$
K({\cal A}) = \sum_{j=1}^{r+q} n_j (r+q - j) - \sum_{j < k}^{r+q} n_j n_k.
$$
Noting
$$
\sum_{j < k}^{r+q} n_j n_k = {1 \over 2} \Big ( \sum_{j=1}^{r+q} n_j \Big )^2 -
 {1 \over 2} \sum_{j=1}^{r+q} n_j = {1 \over 2} ( q^2 - q),
$$
we see that (\ref{L1f}) is equivalent to showing
\begin{equation}\label{L2}
\sum_{\cal A} e^{- 2 \pi i \sum_{j=1}^{r+q} n_j (r+q-j)/(r+1) } = 0.
\end{equation}

Now, the LHS of (\ref{L2}) is equal to the coefficient of $z^q$ in
\begin{equation}\label{L3}
F(z) := \sum_{n_1,\dots,n_{r+q} = 0,1}
e^{- 2 \pi i \sum_{j=1}^{r+q} n_j (r + q - j)/(r+1) }
z^{\sum_{j=1}^{r+q} n_j }.
\end{equation}
The multiple sum factorizes into a product of $r+q$ single sums, giving
the factorization formula
$$
F(z) = \prod_{j=1}^{r+q} (1 + z e^{- 2 \pi i (j-1)/(r+1)}) =
(1 - (-z)^{r+1}) \prod_{j=1}^{q-1} (1 + z e^{-2 \pi i (j-1)/(r+1)}),
$$
where in obtaining the second equality use has been made of the simple identity
$\prod_{l=1}^N(1 - z e^{2 \pi i (l-1)/N}) = 1 - z^N$. We read off from this
that for some $c_1,\dots,c_{q-1}$,
$$
F(z) = 1 + c_1 z + \cdots + c_{q-1} z^{q-1} +
(-1)^r z^{r+1} + (-1)^r c_1 z^{r+2} + \cdots + (-1)^r c_{q-1} z^{r+q}.
$$
In particular (recalling that $1 \le q \le r$) there are no  terms proportional to
$z^q$, and so (\ref{L2}) must hold.
\hfill $\square$

\subsection{Proof of Proposition \ref{pr1}}
As written (\ref{L1}) is only defined for real $\{a_p\}$, ordered
so that $a_1 > \cdots > a_n$. It can be defined for general complex numbers by
analytic continuation. This requires first replacing the absolute value in
(\ref{L1}) according to
$$
| \lambda_j - a_p | = \left \{ \begin{array}{ll}
\lambda_j - a_p, & {\rm for } \: \lambda_j > a_p \\
a_p - \lambda_j, &  {\rm for } \: a_p > \lambda_j,
\end{array} \right.
$$
so that the integrand consists entirely of power functions. According to
Cauchy's theorem, the contours of integration can now be deformed into the
complex plane, provided no contour crosses a branch cut of the power
functions. With the latter specified by $z^\alpha := |z|^\alpha
e^{i \alpha {\rm arg} \, z}$, $- \pi < {\rm arg} \, z \le \pi$, the branch
cut is on the negative real axis of the complex $z$-plane. Moving the position
of the $a_p$'s into the complex plane in this setting gives the analytic
continuation of $L_{r,n}(\{a_p\})$.
 
\begin{figure}[th]
\epsfxsize=10cm
\centerline{\epsfbox{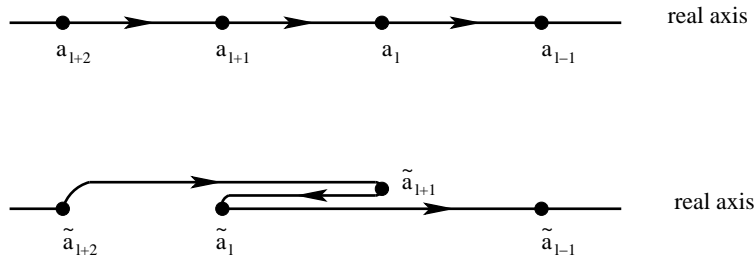}}
\caption{\label{FVfig.1} The contours from $a_{l+2}$ to $a_{l+1}$, $a_{l+1}$ to
$a_l$, and $a_l$ to $a_{l-1}$ are deformed to the contours joining
the corresponding tilded variables. Our interest is in the limit that
$\tilde{a}_j = a_j$ $(j \ne l,l+1)$, $\tilde{a}_l = a_{l+1}$,
$\tilde{a}_{l+1} = a_l$ and all contours in the second diagram run
along the real axis. In the case $l=n-1$ the contour from
$\tilde{a}_{l+2}$ to $\tilde{a}_{l+1}$ is to be deleted, while in the case
$l=1$ the contour from $\tilde{a}_l$ to $\tilde{a}_{l-1}$ is to be deleted.
} 
\end{figure}

Our interest is in this analytic continuation when $a_l$ and $a_{l+1}$ swap places on 
the real axis for general $l=1,2,\dots,n-1$. This analytic continuation corresponds
to the limit that the contours tend to the real line in the second configuration of
Figure \ref{FVfig.1}, and the endpoints $\{\tilde{a}_j\}$ are appropriately related
to the endpoints $\{a_j\}$.

To study the integrand of (\ref{L1}) in the case of the second configuration,
for notational convenience set $\lambda_{(j-1)r + \nu} = \lambda_j^{(\nu)}$
$(\nu = 1,\dots,r)$, which are to be referred to as
species $j$. Then for the first configuration
$$
a_{j+1} < \lambda_j^{(r)} <  \lambda_j^{(r-1)} < \cdots < \lambda_j^{(1)} < a_j
\qquad (j=1,\dots,n-1),
$$
while the integrand itself reads
\begin{eqnarray}\label{L1c}
&& \Big ( \prod_{j=1}^n \prod_{1 \le \nu \le \mu \le r} (\lambda_j^{(\nu)} - \lambda_j^{(\mu)} )^{2/(r+1)}
\Big )
\Big ( \prod_{1 \le j < k \le n} \prod_{\nu,\mu=1}^r ( \lambda_j^{(\nu)} - \lambda_k^{(\mu)})^{2/(r+1)}
\Big )
\nonumber \\
&& \qquad \times \Big ( \prod_{\nu=1}^r \prod_{j=1}^n \prod_{p=1}^n | \lambda_j^{(\nu)} - a_p|^{s_p - 1}
\Big ) =: I_{r,n}(\{\lambda_j^{(\nu)}\};\{a_p\}).
\end{eqnarray}
In the second configuration let the integration variables of (\ref{L1c}) be tilded
so that $\lambda_j^{(\nu)} \mapsto \tilde{\lambda}_j^{(\nu)}$. On each contour from
$\tilde{a}_j$ to $\tilde{a}_{j+1}$ the integration variables $\{\tilde{\lambda}_j^{(\nu)}\}_{\nu=1,\dots,r}$
are ordered so that $\tilde{\lambda}_j^{(1)} \succ \tilde{\lambda}_j^{(2)} \succ \cdots \succ  
\tilde{\lambda}_j^{(r)}$,
where the ordering $\succ$ is the descending order induced by the direction of the contour.
In the limit that all the contours run along the real axis, this ordering corresponds to
\begin{eqnarray}\label{m.3.0}
&& \tilde{a}_j > \tilde{\lambda}_j^{(1)} >  \tilde{\lambda}_j^{(2)} > \cdots >
 \tilde{\lambda}_j^{(r)} > \tilde{a}_{j+1} \qquad j=1,\dots,n-1 \: \: (j \ne l), \nonumber \\
&&  \tilde{a}_{l+1} > \tilde{\lambda}_l^{(r)} > \tilde{\lambda}_l^{(r-1)} > \cdots
>  \tilde{\lambda}_l^{(1)} >   \tilde{a}_{l}.
\end{eqnarray}
Furthermore, we see from Figure \ref{FVfig.1} that for some $p=0,\dots,r$, 
$q=0,\dots,r$,
\begin{eqnarray}
&& \tilde{a}_{l+1} > \tilde{\lambda}_{l+1}^{(1)} > \tilde{\lambda}_{l+1}^{(2)} > \cdots
 \tilde{\lambda}_{l+1}^{(p)} > \tilde{a}_l \label{m.3.1} \\
&& \tilde{a}_{l+1} > \tilde{\lambda}_{l-1}^{(r-q+1)} > \tilde{\lambda}_{l-1}^{(r-q+2)} > \cdots
 > \tilde{\lambda}_{l-1}^{(r)} > \tilde{a}_l \label{m.3.2}
\end{eqnarray}
(note that these configurations are empty if $p=0$, $q=0$ respectively). In other words,
between $\tilde{a}_l$ and $\tilde{a}_{l+1}$ there are $r$ coordinates of species $l$,
$p$ of species $l+1$ and $q$ of species $l-1$.

A crucial feature of the contour integrals is that only configurations with
$p=q=0$ contribute, due to cancellation effects for $p$ and/or $q$ non-zero. To
quantify the latter, consider first the case that $p=0$ while $q \ge 1$, and
suppose that to begin the $r$ species $l$ variables are to the left of the $q$
species $l-1$ variables in the interval $(\tilde{a}_l, \tilde{a}_{l+1})$. We see
from (\ref{L1c}) that interchanging the position of coordinates corresponding to
different species does not change the magnitude of the integrand but it does change
the phase, with each interchange of a species $l-1$ and left neighbouring species
$l$ contributing $e^{-2 \pi i/(r+1)}$. Hence for a general ordering of the $r$ species
$l$ variables and $q$ species $l-1$ variables amongst a given set of $(r+q)$ positions
in $(\tilde{a}_l, \tilde{a}_{l+1})$ the phase is given by
\begin{equation}\label{t.3.1}
e^{-2 \pi i K({\cal A})/(r+1)}.
\end{equation}
Here $K({\cal A})$ is as in Lemma \ref{Lem1} with the 0's corresponding to
species $l$ and the 1's to species $l-1$. But Lemma \ref{Lem1} tells us that
if we sum (\ref{t.3.1}) over all arrangements we get zero, which is the claimed
cancellation effect in this case.

Essentially the same argument, making use of Lemma \ref{Lem1} with the role of the
0's and 1's interchanged, gives  cancellation of the contribution to the contour
integrals from configurations with $q=0$, $p \ge 1$. It remains to consider the cases
$p,q \ge 1$. In such cases, with the positions of the species $l-1$ coordinates fixed
(we could just as well fix the position of the species $l+1$ coordinates), we see that
the contribution to the phase of each such coordinate is equal to $e^{-2 \pi i l^*/(r+1)}$,
where $l^*$ is the number of both species $l$, $l+1$ to its left, and in particular
is independent of their ordering. But we know that summing over this latter ordering
gives the cancellation (\ref{L1}), so in all cases there is no contribution from
non-empty configurations (\ref{m.3.1}) and (\ref{m.3.2}).

As a consequence of both (\ref{m.3.1}) and (\ref{m.3.2}) having to be empty for
a non-zero contribution to the contour integral, it follows that (\ref{m.3.0}) can
be supplemented by the requirements that
\begin{eqnarray*}
&& \tilde{a}_l > \tilde{\lambda}_{l+1}^{(1)} >  \tilde{\lambda}_{l+1}^{(2)} > \cdots
 \tilde{\lambda}_{l+1}^{(r)} > \tilde{a}_{l+2} \\
&& \tilde{a}_{l-1} > \tilde{\lambda}_{l-1}^{(1)} >  \tilde{\lambda}_{l-1}^{(2)} > \cdots
 \tilde{\lambda}_{l-1}^{(r)} > \tilde{a}_{l+1}.
\end{eqnarray*}
Up to a phase, this contour integral is precisely (\ref{L1}) with the position of
$a_l$ and $a_{l+1}$ interchanged, and correspondingly $s_l$ and $s_{l+1}$ interchanged.
The phase is straightforward to calculate, giving as the final result (\ref{Las}).
\hfill $\square$

\subsection{Proof of Theorem 1 for (\ref{d3})}
As remarked below (\ref{R1}), we must show that $L_{r,n}(\{a_p\})$ is
proportional to $R_{r,n}(\{a_p\})$, and then determine the proportionality.
For the former task, our strategy is to show that $L_{r,n}(\{a_p\})$ factorizes
into a term singular in $\{a_p\}$, and a term analytic in $\{a_p\}$. The singular
factor is precisely $R_{r,n}(\{a_p\})$, while a scaling argument shows that
the analytic factor must be a constant.
Intermediate working relating
to the singular terms allows the proportionality to be determined.

Consider $L_{r,n}(\{a_p\})$ as an analytic function of $a_1$ in the
appropriately cut complex $a_1$-plane. Singularites occur as $a_1$ approach any of
$a_2,\dots,a_n$. The singular behaviour as $a_1$ approaches $a_2$ can be
determined directly from (\ref{L1}). Thus, as $a_1 \to a_2$ the integral over
species 1 effectively factorizes from the integral over the other species, showing
\begin{eqnarray}\label{p6}
&& L_{r,n}(\{a_p\}) = \prod_{p=3}^n (a_2 - a_p)^{r (s_p - 1)} I_r(a_1,a_2) \nonumber \\
&& \qquad \times
L_{r,n-1}(\{a_p\}_{p=2,\dots,n}) |_{s_1 \mapsto s_1 + s_2 + 2/(r+1) - 1}
F(a_1 - a_2; \{ a_p\}_{p=2,\dots,n} )
\end{eqnarray}
where $F(z; \{ a_p\}_{p=2,\dots,n} )$ is analytic about $z=0$ and equal to unity
at $z=0$, and
\begin{eqnarray}\label{L2e}
&& I_{r}(a_1,a_2) := \int_{a_1 > \lambda_1^{(1)} > \cdots > \lambda_1^{(r)} > a_2}
d \lambda_1^{(1)} \cdots d \lambda_1^{(r)} \, \nonumber \\
&& \qquad \times
\prod_{1 \le \nu \le \mu \le r} (\lambda_1^{(\nu)} - \lambda_1^{(\mu)})^{2/(r+1)}
\prod_{\nu = 1}^r (a_1 - \lambda_1^{(\nu)})^{s_1 - 1} ( \lambda_1^{(\nu)} - a_2)^{s_2 - 1}.
\end{eqnarray}
Thus the singular behaviour is determined by the singular behaviour of $I_r(a_1,a_2)$.
This in turn is revealed by a simple scaling of the integrand, which shows
\begin{equation}\label{L3s}
I_{r}(a_1,a_2) = (a_1 - a_2)^{r(r-1)/(r+1) + r(s_1 + s_2 - 1)}
{1 \over r!} S_r(s_1-1,s_2,1/(r+1))
\end{equation}
where $S_n(\lambda_1,\lambda_2,\lambda)$ denotes the Selberg integral (\ref{3.1}).

For the singular behaviour as $a_1$ approaches $a_k$ $(k \ne 2)$, we make use of
Proposition \ref{pr1} which says that up to a phase the function of $\{a_p\}_{p=1,\dots,n}$
obtained from $L_{r,n}(\{a_p\})$ by analytic continuation is symmetric in
$\{(a_p,s_p)\}$. Hence as a function of $a_1$ it must be that
\begin{equation}\label{Rag}
L_{r,n}(\{a_p\}) = \prod_{k=2}^n (a_j - a_k)^{r(r-1)/(r+1) + r(s_j + s_k - 1)}
\tilde{F}(a_1;\{a_p\}_{p=2,\dots,n}),
\end{equation}
where $\tilde{F}$ is analytic in $a_1$. Further, repeating the argument with
$L_{r,n}(\{a_p\})$ regarded as a function of $a_2,\dots,a_n$ in turn shows
\begin{equation}\label{RaLu}
L_{r,n}(\{a_p\}) =  R_{r,n}(\{a_p\}) G(\{a_p\})
\end{equation}
where $G$ is analytic in $\{a_p\}$ and symmetric in $\{(a_p,s_p)\}$.

It remains to determine $G$. This can be done by considering the scaling
properties of both sides of (\ref{RaLu}) upon the replacements
$\{a_p\} \mapsto \{c a_p\}$, $c > 0$. After changing variables
$\lambda_k \mapsto c \lambda_k$ $(k=1,\dots,n(r-1))$ in (\ref{L1}) we see
\begin{equation}\label{Lc}
L_{r,n}(\{ca_p\}) = c^{r(n-1) + r(n-1)(r(n-1)-1)/(r+1) + r(n-1) \sum_{p=1}^n (s_p - 1)}
L_{r,n}(\{a_p\})
\end{equation}
while we read off from (\ref{R1}) that
\begin{equation}\label{Rc}
R_{r,n}(\{c a_p\}) = \Big ( \prod_{1 \le j < k \le n} c^{r(s_j + s_k - 2/(r+1))} \Big )
R_{r,n}(\{ a_p\}).
\end{equation}
Straightforward simplification shows that the exponents of $c$ in both equations are in fact
equal, and so $L_{r,n}(\{a_p\})$ and $R_{r,n}(\{a_p\})$ are homogeneous of the same degree.
Thus according to (\ref{RaLu}) the function $G(\{a_p\})$ must be homogeneous of degree 0.
Because $G$ is analytic in $\{a_p\}$, this requirement implies that $G$ is actually
independent of $\{a_p\}$, so  $L_{r,n}(\{a_p\})$ is proportional to $R_{r,n}(\{a_p\})$.

The remaining task is to compute the proportionality. For this purpose, note that
because the normalization is independent of the $a_p$'s,
\begin{equation}
\lim_{a_1, \dots,a_n \to a} L_{r,n}(\{a_j\})/ R_{r,n}(\{a_j\}) =
\hat{C}.
\end{equation}
On the other hand, iteration of (\ref{p6}) with the substitution (\ref{L3s})
allows this limit to be computed in
terms of Selberg integrals, giving the result (\ref{d3a}).
\hfill $\square$

\subsection{Outline of the proof of Theorem \ref{tt1} for (\ref{d4})}
The strategy used to establish Theorem \ref{tt1} in the case of (\ref{d3})
requires only minor adjustment to also establish Theorem \ref{tt1} in the
case of (\ref{d4}). Thus with
\begin{equation}\label{1.84}
Q_{r,n}(\{e^{i \theta_p}\}) :=
\int_{\chi_{R_r}} d \psi_1 \cdots d \psi_{rn} \,
 \prod_{1 \le j < k \le rn}
| e^{i \psi_k} - e^{i \psi_j} |^{2/(r+1)}  
\prod_{j=1}^{rn} \prod_{p=1}^n | e^{i \psi_j} - e^{i \theta_p} |^{\alpha_j - 1} 
\end{equation}
and $w_p := e^{i \theta_p}$, the analytic continuation in $\{w_p\}$ is constructed
by rewriting the absolute values in the integrand according to
$|u - v|^2 = (u-v)(1/u - 1/v)$, writing $z_j = e^{i \psi_j}$
$(j=1,\dots,rn))$ and considering (\ref{1.84}) as a contour integral over arcs
of the unit circle in the appropriate complex $z_j$-planes. By then appealing to
Lemma \ref{Lem1} the analogue of Proposition \ref{pr1} can be proved, establishing
that the analytic continuation of $\tilde{L}(\{w_p\})$ is a symmetric function of
$\{(w_p,\alpha_p)\}_{p=1,\dots,n}$.

The next task is to establish the factorization
\begin{equation}\label{QG}
Q_{r,n}(\{w_p\}) = S_{r,n}(\{w_p\}) \tilde{G}(\{w_p\})
\end{equation}
where
\begin{equation}\label{1.85}
S_{r,n}(\{w_p\}) = \prod_{1 \le j < k \le n}
\Big ( \Big ( 1 - {w_j \over w_k} \Big ) \Big (  1 - {w_k \over w_j} \Big )
\Big )^{r((\alpha_j + \alpha_k)/2 - 1/(r+1))}
\end{equation}
while $\tilde{G}(\{w_p\})$ is a symmetric function of $\{(w_p,\alpha_p)\}$ analytic in
$\{w_p\}$.
This is done by first studying (\ref{1.84}) in the limit $w_1 \to  w_2$
to obtain the analogue of (\ref{Rag}), then using the analogue of Proposition \ref{pr1} 
to deduce (\ref{QG}). But both $Q_{r,n}$ and $S_{r,n}$ are homogeneous of
degree zero, and so $\tilde{G}$ must therefore be a constant.

The constant $\tilde{C}$ is independent of the $\theta_i$'s and so can be
computed according to
\begin{equation}\label{2.25}
\lim_{\theta_1,\dots,\theta_{n-1} \to 0}
Q_{r,n}(\{ e^{i \theta_p} \})/ S_{r,n}(\{ e^{i \theta_p} \})  = \tilde{C}.
\end{equation}
Now it follows from 
(\ref{1.84}) that for $\theta_1 \to \theta_2$,
$$
Q_{r,n}(\{w_p\}) \sim \prod_{p=3}^n | e^{i \theta_p} - e^{i \theta_2} |^{r(\alpha_p - 1)}
 I_r(\theta_2,\theta_1) Q_{r,n-1}(\{e^{i \theta_p}\}_{p=2,\dots,n}) \Big |_{\alpha_2 \mapsto \alpha_1+\alpha_2 +
2r/(r+1) - 1}
$$
where $I_r$ is specified by (\ref{L2e}) and evaluated in terms of the
Selberg integral by (\ref{L3s}). Iterating this and recalling $\theta_{n} = 2 \pi$
allows the limit(\ref{2.25}) to be computed in terms of products of Selberg integrals
and a Morris integral, giving (\ref{d3a1}). 
\hfill $\square$

\section{Random matrix consequences of the case $r=1$}
\setcounter{equation}{0}
It has been pointed out in \cite[Section 4.1]{FR02b} that special cases of (\ref{d1}) permit random matrix
interpretations relating to the Jacobi random matrix ensemble, specified by the eigenvalue
PDF
\begin{equation}\label{d5}
{1 \over C} \prod_{l=1}^N x_l^a (1 - x_l)^b \prod_{1 \le j < k \le N} |x_k - x_j|^\beta, \qquad 0 < x_l < 1,
\end{equation}
or equivalently, with the ordering (\ref{f.2}) assumed and it implicit that the support of
$x^a(1-x)^b$ is $[0,1]$, ME${}_{\beta,N}(x^a(1-x)^b)$,
in the cases $\beta = 1,2$ or 4. These cases can be realized in terms of certain 
random matrices with Gaussian
entries (see e.g.~\cite{Fo02}), the entries being real, complex or quaternion real for $\beta = 1,2$ or 4
respectively. The corresponding PDF on the matrices is then invariant under similarity transformation
by orthogonal $(\beta = 1)$, unitary $(\beta = 2)$, unitary symplectic $(\beta = 4)$ matrices, giving
rise to the alternative notations ME${}_{1,N} \mapsto {\rm OE}_N$, ME${}_{2,N} \mapsto {\rm UE}_N$,
ME${}_{4,N} \mapsto {\rm SE}_N$, which are to be used below. Following \cite{FR01}, let us also
change the name of the special decimation procedure D${}_{2}$, calling it instead by
the word even. 
In terms of these notations, it has been proved in
\cite{FR01} that
\begin{eqnarray}
{\rm even} \, {\rm OE}_{2N+1}(x^{(a-1)/2} (1-x)^{(b-1)/2)}) & = & {\rm SE}_N(x^{a+1}(1-x)^{b+1}) \label{6.1} \\
{\rm even} \, {\rm OE}_{2N}((1-x)^{(b-1)/2}) & = & {\rm SE}_N((1-x)^{b+1}) \label{6.2}.
\end{eqnarray}

Consider first the derivation of (\ref{6.1}) from (\ref{d1}). For this set
\begin{equation}\label{6.3}
n=N+2, \, a_1 = 1, \, a_{N+2} = 0, \, s_1 = (b+1)/2, \, s_{N+2} = (a+1)/2, \, s_j = 2 \: \: (j=2,\dots,N+1).
\end{equation}
Noting that then
\begin{eqnarray}\label{7.1}
&&
\prod_{j=2}^{n-1} a_j^{(a-3)/2} (1 - a_j)^{(b-3)/2}
\prod_{1 \le j < k \le n} (a_j - a_k) 
\prod_{1 \le j < k \le n-1} (\lambda_j - \lambda_k) \prod_{j=1}^{n-1} \prod_{p=1}^n|\lambda_j - a_p|^{s_p - 1}
\nonumber \\
&& \qquad \propto {\rm OE}_{2N+1}(x^{(a-1)/2} (1 - x)^{(b-1)/2)})
\end{eqnarray}
while
\begin{equation}\label{7.2}
\prod_{j=2}^{n-1} a_j^{(a-3)/2} (1 - a_j)^{(b-3)/2}
\prod_{1 \le j < k \le n} (a_j - a_k)^{s_j + s_k} \propto {\rm SE}_N(x^{a+1} (1 - x)^{b+1} )
\end{equation}
we see that (\ref{6.1}) follows from (\ref{d1}) by integrating over $\{\lambda_j\}$ and using the
fact that with respect to the latter variables (\ref{d1}) is a PDF and so integrates to unity. To
derive (\ref{6.2}) from (\ref{d1}), set
\begin{equation}\label{7.3}
n=N+1, \, a_1 = 1, \, s_1=(b+1)/2, \, s_j=2 \: (j=2,\dots,N).
\end{equation}
We then have
\begin{eqnarray}\label{7.4}
&&\prod_{j=2}^n (1 - a_j)^{(b-3)/2} \prod_{1 \le j < k \le n} (a_j - a_k)
\prod_{1 \le j < k \le n - 1} (\lambda_j - \lambda_k)
\prod_{j=1}^{n-1} \prod_{p=1}^n | \lambda_j - a_p|^{s_p - 1}  
\nonumber \\
&& \qquad
\propto {\rm OE}_{2N}((1-x)^{(b-1)/2})
\end{eqnarray}
while
\begin{equation}\label{7.5}
\prod_{j=2}^n (1 - a_j)^{(b-3)/2} \prod_{1 \le j < k \le n} (a_j - a_k)^{s_j+s_k} \propto {\rm SE}_N
((1-x)^{b+1})
\end{equation}
and so (\ref{6.2}) is also seen as a consequence of (\ref{d1}) being a PDF and so integrating to unity.

The inter-relations (\ref{6.1}) and (\ref{6.2}) have companions, which in \cite{FR01} were proved
in fact to be equivalent to the originals. These state that
\begin{equation}
{\rm even} \Big ( {\rm OE}_N(x^{(a-1)/2}(1-x)^{(b-1)/2)} \cup 
{\rm OE}_{N+1}(x^{(a-1)/2}(1-x)^{(b-1)/2)}) \Big )  =  {\rm UE}_N(x^a(1-x^b))
\label{8.1}
\end{equation}
\begin{equation} 
 {\rm even} \Big ( {\rm OE}_N((1-x)^{(b-1)/2)} \cup {\rm OE}_{N}(1-x)^{(b-1)/2)} \Big )  =  {\rm UE}_N((1-x^b)).
\label{8.2}
\end{equation}
In general, for two matrix ensembles ME${}_{\beta_1,n}$ and 
ME${}_{\beta_2,m}$, the operation ME${}_{\beta_1,n} \cup$ME${}_{\beta_2,m}$ denotes the
ensemble of $(n+m)$ eigenvalues formed by sampling independently from 
ME${}_{\beta_1,n}$, ME${}_{\beta_2,m}$,
superimposing the resulting eigenvalue sequences, and labelling from right to left.
We know from \cite{FR01} that the eigenvalue PDF of OE${}_n (f)\cup$OE${}_{n+1}(f)$ is proportional to
\begin{equation}\label{9.1}
\prod_{l=1}^{2n+1} f(x_l) \prod_{1 \le j < k \le n+1} (x_{2j-1} - x_{2k-1}) \prod_{1 \le j < k \le n} (x_{2j} - x_{2k})
\end{equation}
while the PDF of OE${}_n(f) \cup$OE${}_{n}(f)$ is proportional to
\begin{equation}\label{9.2}
\prod_{l=1}^{2n} f(x_l) \prod_{1 \le j < k \le n} (x_{2j-1} - x_{2k-1}) (x_{2j} - x_{2k}).
\end{equation}

Given these facts, the inter-relations (\ref{8.1}) and (\ref{8.2}) can be understood as corollaries of (\ref{d1}).
Explicitly, to deduce (\ref{8.1}), choose the parameters as in (\ref{6.3}) except that $s_j=1$, $(j=1,\dots,N)$.
Then, recalling (\ref{9.1}), we have that the LHS of (\ref{7.1}) is proportional to the LHS of (\ref{8.1}),
while the LHS of (\ref{7.2}) is proportional to the RHS of (\ref{8.1}). Similarly, to deduce (\ref{8.2})
from (\ref{d1}), choose the parameters as in (\ref{7.3}) but with $s_j=1$ $(j=2,\dots,N)$. Recalling
(\ref{9.2}), we then have that the LHS of (\ref{7.4}) is proportional to the LHS of (\ref{8.2}),
while the LHS of (\ref{7.5}) is proportional to the RHS of (\ref{8.2}).

The Jacobi $\beta$-ensemble (\ref{d5}) permits two well known limiting cases. Thus, with the replacements
$x_j \mapsto x_j/b$ and taking the limit $b \to \infty$, one obtains the eigenvalue PDF for the
Laguerre $\beta$-ensemble,
\begin{equation}\label{L1x}
{1 \over C} \prod_{l=1}^N x_l^a e^{- x_l} \prod_{1 \le j < k \le N} |x_k - x_j|^\beta, \qquad 0 < x_l < \infty.
\end{equation}
Also, with $a=b$ and $x_j \mapsto {1 \over 2} (1 - x_j/\sqrt{2 b})$, taking the limit $b \to \infty$
one obtains the eigenvalue PDF for the Gaussian $\beta$-ensemble
\begin{equation}\label{G1}
{1 \over C} \prod_{l=1}^N e^{- x_l^2/2} \prod_{1 \le j < k \le N} |x_k - x_j|^\beta.
\end{equation}
The inter-relations (\ref{6.1}), (\ref{6.2}) and (\ref{8.1}), (\ref{8.2}) permit these same limiting
cases. 
For the Laguerre limit we read off from these that
\begin{eqnarray}\label{G2}
&& {\rm even} \, {\rm OE}_{2N+1}(x^{(a-1)/2} e^{-x/2}) = {\rm SE}_N (x^{a+1} e^{-x}) \nonumber \\
&& {\rm even} \, {\rm OE}_{2N}(e^{-x/2}) = {\rm SE}_N (e^{-x}) \nonumber \\
&& {\rm even} \, ({\rm OE}_N (x^{(a-1)/2} e^{-x/2}) \cup {\rm OE}_{N+1} (x^{(a-1)/2} e^{-x/2})) =
{\rm UE}_N (x^a e^{-x} ) \nonumber \\
&& {\rm even} \, ({\rm OE}_N ( e^{-x/2}) \cup {\rm OE}_{N+1} ( e^{-x/2})) =
{\rm UE}_N (e^{-x} ),
\end{eqnarray}
while only (\ref{6.1}) and (\ref{8.1}) permit Gaussian limits, which read
\begin{eqnarray}\label{G3}
&& {\rm even} \,  {\rm OE}_{2N+1}(e^{- x^2/2}) = {\rm SE}_N (e^{- x^2}) \nonumber \\
&& {\rm even} \, ( {\rm OE}_{N}(e^{- x^2/2}) \cup {\rm OE}_{N+1}(e^{- x^2/2}) ) = {\rm UE}_N(e^{-x^2}).
\end{eqnarray}
These were first obtained in \cite{FR01} in the context of  classification 
theorems relating to the form of $f(x)$ in OE${}_N(f)$ 
which permits such identities. We remark too that as observed in \cite{FR02b}, 
the Dixon-Anderson PDF (\ref{d1})
permits analogous Laguerre and Gaussian limits, and these have (\ref{G2}) and (\ref{G3}) as
corollaries.

We now turn our attention to the revision of some consequences in random matrix theory relating to the
conditional PDF (\ref{d2}) \cite{FR05}. For this we recall that the circular ensembles COE${}_n$, CUE${}_n$,
CSE${}_n$ of unitary random matrices have their eigenvalue PDF proportional to
\begin{equation}\label{11.2}
\prod_{1 \le j < k \le n} |e^{i \theta_k} - e^{i \theta_j} |^\beta
\end{equation}
for $\beta = 1,2,4$ respectively. Further, we require knowledge of the fact that the PDF of
COE${}_n \cup$COE${}_n$ is proportional to \cite{Gu62}
\begin{equation}\label{12.1}
\prod_{1 \le j < k \le n} \sin (\theta_{2k} - \theta_{2j})/2 \, \sin (\theta_{2k-1} - \theta_{2j-1})/2
\end{equation}
with $\{ \theta_{2j} \}$ and $\{\theta_{2j-1} \}$ interlaced according to 
$\theta_{2j-1} < \theta_{2j} <
\theta_{2j+1}$ ($j=1,\dots,n$), $\theta_{2n+1} :=2 \pi$.

In (\ref{d2}) regard $\theta_n$ as a variable with
\begin{equation}\label{10.1}
\psi_n < \theta_n, \qquad \theta_n {\rm mod} \, 2 \pi < \psi_1.
\end{equation}
Let $\tilde{R}$ denote the interlaced region (\ref{2.1}) supplemented by (\ref{10.1}). Then, from the fact
that (\ref{d2}) is a PDF in $\{\psi_j\}_{j=1,\dots,n}$ we have
\begin{eqnarray}\label{11.1}
&&\int_{\tilde{R}} d \psi_1 \cdots d \psi_n \,
\prod_{1 \le j < k \le n} | e^{i \theta_k} - e^{i \theta_j} | | e^{i \psi_k} - e^{i \psi_j} |
\prod_{j=1}^n \prod_{l=1}^n | e^{i \theta_j} - e^{i \psi_l} |^{\alpha_j - 1} \nonumber \\
&& \qquad \propto  \prod_{1 \le j < k \le n} | e^{i \theta_k} - e^{i \theta_j} |^{\alpha_j + \alpha_k}.
\end{eqnarray}
Setting $\alpha_j = 2$ $(j=1,\dots,n)$ we obtain the inter-relation between circular ensembles
\cite{DM63}
\begin{equation}\label{3.23}
{\rm alt} \, {\rm COE}_{2n} = {\rm CSE}_n,
\end{equation}
where the operation alt refers to the distribution of every alternate (second) eigenvalue as ordered
around the circle. Further, after making use of (\ref{12.1}), setting $\alpha_j = 1$ $(j=1,\dots,n)$
we obtain the companion inter-relation
\begin{equation}
{\rm alt} \, ({\rm COE}_n \cup {\rm COE}_n) = {\rm CUE}_n.
\end{equation}

An ensemble which generalizes (\ref{11.2}) is the circular Jacobi $\beta$-ensemble
(\ref{f.3}), which in the special cases of $\beta=1,2,4$ is to be referred to as
COE${}_n^b$, CUE${}_n^b$, CSE${}_n^b$
respectively. It follows from (\ref{d2}) itself (and thus $e^{i \theta_{n}}$ therein
equal to unity) that
\begin{equation}\label{3.25}
{\rm even} \, {\rm COE}_{2N+1}^b = {\rm CSE}_N^{2b+2}, \qquad
{\rm even} ( {\rm COE}_{N+1}^b \cup {\rm COE}_N^b) = {\rm CUE}_N^{2b+1}.
\end{equation}
The first of these contains (\ref{3.23}) as a special case. To see this, note from
(\ref{f.3}) that in general 
\begin{equation}\label{CE}
{\rm CE}_{\beta,N}^\beta \propto {\rm CE}_{\beta,N+1}^0 |_{\theta_{N+1} = 0}.
\end{equation}
Setting $b=1$ in the first formula of (\ref{3.25}) is then seen to reclaim (\ref{3.23})
with $n=N+1$.

\section{Random matrix consequences for general $r$}
\setcounter{equation}{0}
In this section Theorem \ref{tt2} will be established.
Consider the integrand (\ref{L1}) in the case
\begin{equation}\label{2pa}
n=N+2, \: a_1=1, \: a_{N+2} = 0, \:
s_1 = b+1, \: s_{N+2} = a+1, \: s_j = 1 + 2/(r+1) \quad (j=2,\dots,N+1)
\end{equation}
(cf.~(\ref{6.3})).
After multiplication by a suitable function of $\{a_p\}$ we see that a matrix ensemble
PDF of the form (\ref{f.1}) results. Explicitly one has
\begin{eqnarray}\label{6.1f}
&& \prod_{j=2}^{n-1} a_j^{a - 2/(r+1)} (1 - a_j)^{b - 2/(r+1)}
\prod_{1 \le j < k \le n} (a_j - a_k)^{2/(r+1)}
\prod_{1 \le j < k \le r(n-1)} (\lambda_j - \lambda_k)^{2/(r+1)} \nonumber \\
&& \qquad \times \prod_{j=1}^{r(n-1)} \prod_{p=1}^n | \lambda_j - a_p|^{s_p-1}
\propto {\rm ME}_{2/(r+1), (r+1) N + r} (x^a (1 - x)^b).
\end{eqnarray}
Multiplication of (\ref{R1}) by this same factor gives
\begin{eqnarray}\label{6.2f}
&&
\prod_{j=2}^{n-1} a_j^{a - 2/(r+1)} (1 - a_j)^{b - 2/(r+1)}
\prod_{1 \le j < k \le n} (a_j - a_k)^{r(s_j + s_k) - 2(r-1)/(r+1)} \nonumber \\
&& \qquad \propto {\rm ME}_{2(r+1), N} ( x^{(r+1)a + 2r} (1 - x)^{(r+1)b+2r} ).
\end{eqnarray}

The fact, as implied by Theorem \ref{tt1} as it relates to (\ref{d3}), that (\ref{L1})
and (\ref{R1}) are proportional, tells us that integrating (\ref{6.1f}) over the variables
$\{\lambda_j \}$, (\ref{6.2f}) results. Now, by the ordering implied by (\ref{1.6a}), and the
fact that $a_1$ is fixed, integrating over the variables $\{\lambda_j\}$ gives the joint
marginal distribution of the variables in positions $(r+1), 2(r+1), 3(r+1), \dots$ as
counted in a descending order. Thus the integration in (\ref{L1}) is the operation
D${}_{r+1}$, and so the equality between integrating (\ref{6.1}) over
$\{\lambda_j\}$ and (\ref{6.2}) can be expressed as the first of the relations in
(\ref{tt3}). 

To derive the second relation, set
$$
n = N+1, \: a_1 = 1, \: s_1 = b+1, \: s_j = 1+2/(r+1) \quad (j=2,\dots,N)
$$
(cf.~(\ref{7.3}))
in the integrand of (\ref{L1}). The integrand can then be multiplied by a suitable function
of $\{a_p\}$ to obtain an eigenvalue PDF of the form (\ref{f.1}),
\begin{eqnarray*}
&&  \prod_{j=2}^{n}  (1 - a_j)^{b - 2/(r+1)} \prod_{1 \le j < k \le n} (a_j - a_k)^{2/(r+1)}
\prod_{1 \le j < k \le r(n-1)} (\lambda_j - \lambda_k)^{2/(r+1)} \nonumber \\
&& \qquad \times \prod_{j=1}^{r(n-1)} \prod_{p=1}^n | \lambda_j - a_p|^{s_p-1}
\propto {\rm ME}_{2/(r+1), (r+1) N } ((1 - x)^b).
\end{eqnarray*}
Integration over $\{\lambda_j\}$ corresponds to the decimation procedure 
D${}_{r+1}$,
while multiplication of (\ref{R1}) by this same function of $\{a_p\}$ gives
\begin{eqnarray*}
&&
\prod_{j=2}^{n} a_j^{b - 2/(r+1)} (1 - a_j)^{b - 2/(r+1)}
\prod_{1 \le j < k \le n} (a_j - a_k)^{r(s_j + s_k) - 2(r-1)/(r+1)} \nonumber \\
&& \qquad \propto {\rm ME}_{2(r+1), N} (  (1 - x)^{(r+1)b+2r} ),
\end{eqnarray*}
and the sought identity results.

Changing variables $x_j \mapsto x_j/b$ in the first and second relations and
taking $b \to \infty$ gives the third and fourth relations. This is the Laguerre
limit discussed in the sentence including (\ref{L1x}). In the first identity, setting
$a=b$, changing variables $x_j \mapsto {1 \over 2} (1 - x_j)/\sqrt{2b}$, and then taking
$b \to \infty$ to obtain the Gaussian limit as discussed in the sentence including
(\ref{G2}) gives the fifth relation.

The sixth relation is a consequence of Theorem \ref{tt1} as it applies to (\ref{d4}), which
tells us that (\ref{1.84}) is proportional to (\ref{1.85}) with $w_p = e^{i \theta_p}$ in the
latter. For this set
$$
n = N+1, \: \alpha_n = b+1, \: \alpha_j = 1+ 2/(r+1) \quad (j=1,\dots,n-1).
$$
The final relation is obtained from this by setting $b=0$, replacing $N$ by $N-1$, and
recalling (\ref{CE}). 

\section{Gap and spacing probabilities}
\setcounter{equation}{0}
\subsection{A class of gap probabilities}
It has been revised in Section 3 that the Dixon-Anderson density (\ref{d1}) with $s_j = 1$
implies the decimation identities for superimposed ensembles (\ref{8.1}),
(\ref{8.2}). These results rely on the eigenvalue PDF of ${\rm OE}_n(f) \cup
{\rm OE}_{n+1}(f)$ being proportional to (\ref{9.1}) and ${\rm OE}_n(f) \cup
{\rm OE}_{n}(f)$ being proportional to (\ref{9.2}). No generalization of these latter facts for
ensembles of the form (\ref{f.1}) is known. Instead let us make note of a further interpretation
of the Dixon-Anderson integral in the case $s_j = 1$, or more explicitly in the case of the
parameters (\ref{6.3}) with the final condition replaced by $s_j=1$ $(j=2,\dots,N+1)$, which does
permit a generalization.

In the latter circumstance, one has that
\begin{eqnarray}\label{w.2}
&& \int_A d \lambda_1 \cdots d \lambda_{N+1} \, \prod_{j=1}^{N+1} \lambda_j^a (1 - \lambda_j)^b
\prod_{1 \le j < k \le N+1} (\lambda_j - \lambda_k) \nonumber \\
&& \qquad = {\Gamma(a+1) \Gamma(b+1) \over \Gamma(a+b+N+2) }
\prod_{j=1}^N a_j^{a+1} (1 - a_j)^{b+1} \prod_{1 \le j < k \le N} (a_j - a_k)
\end{eqnarray}
where $A$ denotes the region
\begin{equation}\label{w.3}
1 > \lambda_1 > a_1 > \lambda_2 > a_2 > \cdots > \lambda_{N+1} > 0.
\end{equation}
Let $1/C_{N+1}$ be the normalization such that the integrand of (\ref{w.2}) corresponds to the
eigenvalue PDF OE${}_{N+1}(x^a(1-x)^b)$. Then with $A$ denoting the inequalities (\ref{w.3})
we see that (\ref{w.2}) has the interpretation as the evaluation of the
probability that there is exactly one
eigenvalue in each interval $[a_{j+1},a_j]$ $(j=0,\dots,N; a_0=1, a_{N+1} = 1$), giving
the result
$$
{\rm Pr}(A; {\rm OE}_{N+1}(x^a(1-x)^b)) =
{1 \over C_{N+1} } {\Gamma(a+1) \Gamma(b+1) \over \Gamma(a+b+N+2) }
\prod_{j=1}^N a_j^{a+1} (1 - a_j)^{b+1} \prod_{1 \le j < k \le N} (a_j - a_k).
$$
Similarly, replacing the condition $s_j=2$ in (\ref{6.3}) by $s_j=1$ we can obtain an
analogous formula relating to the probability of the inequalities (\ref{1.6a})
holding in the ensemble ME${}_{2/(r+1),(r+1)N+r}(x^a(1-x)^b)$. However
such formulas appear to be of limited interest, so we refrain from
pursuing this further.

\subsection{Spacing probabilities}\label{ssp}
The inter-relations of Theorem \ref{tt2} have some immediate consequences for the PDF of
ordered eigenvalues in the corresponding ensembles. To state these,
let $p^{\rm max}(k;s;{\rm ME}_{\beta,N})$ denote the PDF for the event that in the matrix
ensemble ME${}_{\beta,N}$ there is an eigenvalue at $s$, and exactly $k$
eigenvalues to the right of $s$. Thus this is the PDF for the $(k+1)$-st eigenvalue
as labelled from the largest. Similarly, let $p^{\rm min}(k;s;{\rm ME}_{\beta,N})$ 
denote the PDF for the event that in the matrix
ensemble ME${}_{\beta,N}$ there is an eigenvalue at $s$, and exactly $k$
eigenvalues between 0 and $s$. In the case that the support of the spectrum
is restricted to $x>0$, this corresponds to the PDF for the $(k+1)$-st eigenvalue
as labelled from the smallest. We then read off from the fourth, third and sixth relations
in (\ref{tt3}) the following result (analogous results hold for the other relations;
however these three are representative of all situations). 

\begin{cor}\label{cv1}
One has
\begin{eqnarray}\label{pmax}
p^{\rm max}((r+1)k + r; s; {\rm ME}_{2/(r+1), (r+1)N+r}(e^{-x})) & = &
p^{\rm max}(k;s;{\rm ME}_{2(r+1),N}(e^{-(r+1)x})) \nonumber \\
p^{\rm min}((r+1)k + r; s; {\rm ME}_{2/(r+1), (r+1)N+r}(x^a e^{-x})) & = &
p^{\rm min}(k;s;{\rm ME}_{2(r+1),N}(x^{(r+1)a + 2r} e^{-(r+1)x})) \nonumber \\
p^{\rm min}(((r+1)k + r; s; {\rm CE}_{2/(r+1), (r+1)N+r}^b & = &
p^{\rm min}(k;s;{\rm CE}_{2(r+1),N}^{(r+1)b + 2r}).
\end{eqnarray}
\end{cor} 
Recalling (\ref{CE}) we see that with $b=2/(r+1)$ the final equation in
(\ref{pmax}) implies
\begin{equation}\label{p00}
p^{\rm spacing}((r+1)k+r;s;{\rm CE}_{2/(r+1), (r+1)N}^0) = 
p^{\rm spacing}(k;s;{\rm CE}_{2(r+1),N}^{0}),
\end{equation}
where $p^{\rm spacing}(k;s;{\rm CE}_{\beta,N})$ is the PDF for the spacing
between eigenvalues which are $(k+1)$-st neighbours.  

The three situations that the results of Corollary \ref{cv1} are representative of
the soft edge, the hard edge, and a spectrum singularity in the bulk. 
The soft edge is the neighbourhood of the
largest eigenvalue, so called because the eigenvalue density has support beyond this
region. The hard edge is the neighbourhood of the smallest eigenvalue, in the
situation that the eigenvalue support is strictly zero for $x<0$. The bulk is the
portion of the spectrum a macroscopic distance from the edges
(all circular ensembles satisfy this requirement), while a spectrum
singularity corresponds to a factor of the form $|x|^\alpha$ (for $x \to 0$) in the
one body weight. 

Each permits a scaling in which the origin is shifted to the
appropriate neighbourhood (this must be done in the case of the soft edge only),
and the eigenvalues are scaled so that the spacing between eigenvalues is
of order unity. For the $N \to \infty$ limit of the ensemble ME${}_{\beta,N}(e^{-cx})$,
the soft edge scaling is given by \cite{Fo93a}
\begin{equation}\label{s1}
x \mapsto {\beta \over 2c} (4N + 2(2N)^{1/3} s_\beta x),
\end{equation}
where $s_\beta > 0$ is an arbitrary length scale. The PDF for the $(k+1)$-st largest eigenvalue
is then given by
\begin{equation}
\lim_{N \to \infty}  {\beta \over c} (2N)^{1/3} s_\beta
p^{\rm max}(k;s;{\rm ME}_{\beta,N}(e^{-cx})) =:
p_\beta^{\rm soft}(k;s).
\end{equation}
The hard edge scaling of ME${}_{\beta,N}(x^a e^{-cx})$ is given by
\begin{equation}\label{s2}
x \mapsto {\beta \over 2c} {x \over 4 N \tilde{s}_\beta}
\end{equation}
where $\tilde{s}_\beta > 0$ is an arbitrary length scale, and the PDF for the
$(k+1)$-st smallest eigenvalue at the hard edge is thus
\begin{equation}
\lim_{N \to \infty}  {\beta \over 2 c}  {1 \over 4 N \tilde{s}_\beta}
p^{\rm min}(k;s;{\rm ME}_{\beta,N}(x^a e^{-c x})) =: p_\beta^{\rm hard}(k;s;a).
\end{equation}
Finally, in the ensemble CE${}_{\beta,N}^b$ the mean spacing between eigenvalues
is $2 \pi / N$, and we have
\begin{equation}
\lim_{N \to \infty} {2 \pi \over N} p^{\rm min}(k;2 \pi s /N; {\rm CE}_{\beta,N}^b) =:
p_\beta^{\rm bulk, s.s.}(k;s;b)
\end{equation}
for the PDF of the $(k+1)$-st eigenvalue to the right of a spectrum singularity in the bulk 
with unit density.
Taking these limits in the results (\ref{pmax}) gives inter-relations between the scaled
PDFs with $\beta = 2/(r+1)$ and $\beta = 2(r+1)$.

\begin{prop}
Choose the scale in (\ref{s1}) such that $s_{2/(r+1)} = (r+1)^{2/3} s_{2(r+1)}$, and choose the
scale in (\ref{s2}) such that $\tilde{s}_{2/(r+1)} (r+1)^2 = \tilde{s}_{2(r+1)}$. One has
\begin{eqnarray}\label{ksk1}
p_{2/(r+1)}^{\rm soft}((r+1)k+r;s) & = & p_{2(r+1)}^{\rm soft}(k;s) \nonumber \\
p_{2/(r+1)}^{\rm hard}((r+1)k+r;s;a) & = & p_{2(r+1)}^{\rm hard}(k;s;(r+1)a + 2r) \nonumber \\
(r+1) p_{2/(r+1)}^{\rm bulk, s.s.}((r+1)k+r;(r+1)s;b) & = &
p_{2(r+1)}^{\rm bulk, s.s.}(k;s;(r+1)b + 2r).
\end{eqnarray}
\end{prop}
Let $p_\beta^{\rm bulk, sp.}(k;s)$ denote the PDF for a spacing of size $s$ 
between eigenvalues which are $(k+1)$-st neighbours, in the 
bulk of the circular ensemble CE${}_{\beta,N}$ scaled so that the eigenvalue
density is unity in the $N \to \infty$ limit. This corresponds to the third
relation in (\ref{ksk1}) with $b=2/(r+1)$ (recall the discussion leading to (\ref{p00})) 
and so we have
\begin{equation}\label{ksk2}
(r+1) p_{2/(r+1)}^{\rm bulk, sp.}((r+1)k+r;(r+1)s)  = 
p_{2(r+1)}^{\rm bulk, sp.}(k;s).
\end{equation}

As a consistency check on (\ref{ksk2}) it can be shown to be compatible with
the asymptotic form (\ref{10.m2}). First one recalls that in general
$p_\beta^{\rm bulk,sp.}(k;s)$ is related to
$\{E_\beta^{\rm bulk}(n;s) \}_{n=0,1,\dots,k}$ according to the formula
(see e.g.~\cite[Ch.~6]{Fo02})
$$
p_\beta^{\rm bulk,sp.}(k;s) = {d^2 \over d s^2} \sum_{j=0}^k (k-j+1)
E_\beta^{\rm bulk}(j;s).
$$
For large $s$ the term $j=k$ will dominate, and so (\ref{ksk2}) requires that for
large $s$
$$
(r+1) E_{2/(r+1)}^{\rm bulk}((r+1)k;(r+1)s) \: \sim \: E_{2(r+1)}^{\rm bulk}(k;s).
$$
Taking logarithms of both sides and comparing with (\ref{10.m2}) gives 
precise agreement
with the latter. 

\section*{Acknowledgements}
The idea of seeking inter-relations of the type reported on here is
due to B\'alint Vir\'ag, communicated to the author at the
AMS-IMS-SIAM summer research conference on
Random Matrix Theory, Integrable
Systems, and Stochastic
Processes (June, 2007).
Ths work has been supported by the Australian Research Council.


\begin{thebibliography}{10}

\bibitem{An91}
G.W. Anderson.
\newblock A short proof of {Selberg's} generalized beta formula.
\newblock {\em Forum Math.}, 3:415--417, 1991.

\bibitem{BF97a}
T.H. Baker and P.J. Forrester.
\newblock The {Calogero-Sutherland} model and generalized classical
  polynomials.
\newblock {\em Commun. Math. Phys.}, 188:175--216, 1997.

\bibitem{Di05}
A.L. Dixon.
\newblock Generalizations of {L}egendre's formula $k e' - (k - e) k' = {1 \over
  2} \pi$.
\newblock {\em Proc. London Math. Soc.}, 3:206--224, 1905.

\bibitem{DE02}
I.~Dumitriu and A.~Edelman.
\newblock Matrix models for beta ensembles.
\newblock {\em J. Math. Phys.}, 43:5830--5847, 2002.

\bibitem{Dy95}
F.J. Dyson.
\newblock The coulomb fluid and the fifth {Painlev\'e} transcendent.
\newblock In S.-T. Yau, editor, {\em Chen Ning Yang}, page 131. International
  Press, Cambridge MA, 1995.

\bibitem{DM63}
F.J. Dyson and M.L. Mehta.
\newblock Statistical theory of the energy levels of complex systems. {IV}.
\newblock {\em J. Math. Phys.}, 4:701--712, 1963.

\bibitem{ES06}
A.~Edelman and B.D. Sutton.
\newblock From random matrices to stochastic operators.
\newblock math-ph/0607038, 2006.

\bibitem{FS95}
M.~Fogler and B.I. Shklovskii.
\newblock The probability of an eigenvalue fluctuation in an interval of a
  random matrix spectrum.
\newblock {\em Phys. Rev. Lett.}, 74:3312, 1995.

\bibitem{Fo02}
P.J. Forrester.
\newblock Log-gases and {Random} {Matrices}.
\newblock www.ms.unimelb.edu.au/\~{}matpjf/matpjf.html.

\bibitem{Fo93a}
P.J. Forrester.
\newblock The spectrum edge of random matrix ensembles.
\newblock {\em Nucl. Phys. B}, 402:709--728, 1993.

\bibitem{FR01}
P.J. Forrester and E.M. Rains.
\newblock Inter-relationships between orthogonal, unitary and symplectic matrix
  ensembles.
\newblock In P.M. Bleher and A.R. Its, editors, {\em Random matrix models and
  their applications}, volume~40 of {\em Mathematical Sciences Research
  Institute Publications}, pages 171--208. Cambridge University Press, United
  Kingdom, 2001.

\bibitem{FR02b}
P.J. Forrester and E.M. Rains.
\newblock Interpretations of some parameter dependent generalizations of
  classical matrix ensembles.
\newblock {\em Probab. Theory Relat. Fields}, 131:1--61, 2005.

\bibitem{FR05}
P.J. Forrester and E.M. Rains.
\newblock Jacobians and rank 1 perturbations relating to unitary Hessenberg
  matrices.
\newblock {\em Int. Math. Res. Not.}, 2006:48306 (36 pages), 2006.

\bibitem{FW07}
P.J. Forrester and S.O. Warnaar.
\newblock The importance of the {S}elberg integral.
\newblock arXiv:0710.3981v1[math.CA], 2007.

\bibitem{Gu62}
J.~Gunson.
\newblock Proof of a conjecture of {Dyson} in the statistical theory of energy
  levels.
\newblock {\em J. Math. Phys.}, 4:752--753, 1962.

\bibitem{KS06}
R.~Killip and M.~Stoiciu.
\newblock Eigenvalue statistics for {CMV} matrices: from {P}oisson to clock via
  circular beta ensembles.
\newblock math-ph/0608002, 2006.

\bibitem{Ok97}
A.~Okounkov.
\newblock Binomial formula for {Macdonald} polynomials.
\newblock {\em Math. Res. Lett.}, 4:533--553, 1997.

\bibitem{Ra06}
E.M. Rains.
\newblock Limits of elliptic hypergeometric integrals.
\newblock arXiv:math.CA/0607093, 2006.

\bibitem{RRV06}
J.. Ramirez, B.~Rider, and B.~Vir\'ag.
\newblock Beta ensembles, stochastic {A}iry spectrum, and a diffusion.
\newblock arXiv:math/060733[math.PR], 2006.

\bibitem{Se44}
A.~Selberg.
\newblock Bemerkninger om et multipelt integral.
\newblock {\em Norsk. Mat. Tidsskr.}, 24:71--78, 1944.

\bibitem{VV07}
B.~Valko and B.~Vir\'ag.
\newblock Manuscript in preparation, 2007.

\end{thebibliography}

\end{document}